\documentclass[prc,showpacs,preprintnumbers,amsmath,amssymb]{revtex4}
\usepackage{graphicx}

\begin{document}

\title{Weakly bound nuclei and realistic NN interactions}
\author{E. N. E. van Dalen, P. G\"ogelein, and H. M\"uther}
\affiliation{Institut f\"ur Theoretische Physik, \\
Universit\"at T\"ubingen, D-72076 T\"ubingen, Germany}

\begin{abstract}
Nuclei close to the neutron drip line are described employing an interaction
model  which is based on the low-momentum interaction $V_{lowk}$. This effective
two-body  interaction which is determined to reproduce the nucleon-nucleon (NN)
scattering data  at energies below the pion thresh-hold is supplemented by a
density-dependent contact interaction fitted to reproduce the saturation
properties of infinite nuclear matter within the Hartree-Fock approach. It is
demonstrated that corresponding calculations for closed shell-nuclei using this
interaction model reproduce the bulk properties of these nuclei, independent
whether the wave functions are expanded in terms of harmonic oscillator waves or
in a basis of plane waves discretized in a spherical box of appropriate size.
This plane wave basis, however, is more appropriate to describe weakly bound
nuclei and the transition from discrete nuclei to homogeneous matter which is
supposed to occur e.g. in the crust of neutron stars. Properties of exotic
nuclei are studied within a Hartree-Fock plus BCS approximation.

\end{abstract}


\maketitle

\section{Introduction}
The new generation of radioactive beam facilities, e.g. the future GSI facility
FAIR in Germany, the Rare Isotope accelerator planned in the United States of
America or SPIRAL2 at GANIL/France, facilitate the nuclear structure studies
away from the line of $\beta$-stability, especially for the neutron-rich nuclei.
The study of these nuclei is of high interest since they represent important
steps in the nuclear reaction chains for the formation of elements. The weak
binding energies of these nuclei and the corresponding large spatial extensions
of the orbits for the valence nucleons lead to very interesting features like
e.g.  a neutron halo made of several neutrons outside the core. 

These nuclei can furthermore be understood as a first step in the transition
from isolated nuclei to infinite homogeneous matter, which should occur in the
outer crust of a neutron star. This transition is a very good example for a
quantum liquid with a phase transition from the droplet to the homogeneous
phase. This transition is a challenge for theoretical nuclear structure physics
as it incorporates the transition from isolated nuclei via a crystal-like
structure of quasi-nuclei embedded in a sea of neutrons to the phase of homogeneous
baryon matter. These structures have been described by means of Thomas-Fermi
calculations or mean-field calculations in a Wigner-Seitz cell using simple
phenomenological models for the nucleon-nucleon (NN) interaction like the
density-dependent Skyrme 
forces\cite{sk1,sk2,bv81,Oyamatsu:2007,Montani:2004,Goegelein:2007} 
or a relativistic mean-field approach\cite{goegeleina:2008}. 

All these investigations are based on phenomenological interactions, 
which have been adjusted to describe the saturation properties of symmetric
nuclear matter and the structure of stable nuclei with large binding energies
located in the valley of $\beta$-equilibrium. These models and interactions 
have been developed to provide a simple description of the mean field in terms
of local single-particle densities. The predictive power of such simple
phenomenological interactions, however, may be rather limited as these models
have been constructed to fit the data of finite nuclei.

An alternative to this phenomenological approach is the use of so-called
realistic interactions like the charge-dependent Bonn interaction\cite{mach1} 
or one of the Argonne interaction models\cite{arg1}. Such realistic interactions
are based e.g. on the meson exchange model and
adjusted to describe the experimental data of nucleon-nucleon (NN) scattering.
These interactions contain rather strong short range and tensor
components, which make it inevitable to employ non-perturbative approximation
schemes for the solution of the many-body problem for the nuclear hamiltonian
based on such interactions\cite{Muether00}. Nuclear structure calculations using
such interactions have become feasible for very light nuclei and provide a very
good description if an appropriate three-nucleon force has been
added\cite{Pieper:2006}. In the next future such sophisticated calculations 
will not be possible for heavier nuclei or the nuclear structures in the crust
of neutron stars mentioned above.

A possible way out of this problem is to consider an interaction model, which
separates the low-momentum (below a cut-off $\Lambda$) and high-momentum 
components of a realistic NN interaction by means of renormalization
techniques\cite{bogner01,bogner:2005,bogner:2007,bozek:2006}. If the cutoff
$\Lambda$ is chosen around $\Lambda$ = 2 fm$^{-1}$ the resulting low-momentum
interaction $V_{lowk}$ still describes the NN scattering data up to the pion
threshold and turns out to be independent of the underlying realistic 
interaction $V$. Since the high-momentum components, which correspond to the
short-distance behavior, of $V$ have been removed, the resulting $V_{lowk}$
does not produce short-range correlations in a significant way, but can be 
treated within the Hartree-Fock approximation\cite{bozek:2006}. 

Employing $V_{lowk}$ in a calculation of nuclear matter, however, one obtains a
binding energy per nucleon increasing with density in a monotonic way, such that
no saturation point is obtained\cite{bozek:2006,kuckei:2003}. In order to
compensate this deficiency we will supplement $V_{lowk}$ by a simple
density-dependent  contact interaction adjusted to reproduce the empirical
saturation property of symmetric nuclear matter. This density-dependent contact
interaction, which is introduced to obtain the empirical saturation point of
nuclear matter, can also be interpreted as a parameterization of the
three-nucleon force, which, as discussed above, seems to be necessary for a
microscopic description of nuclear structure data based on realistic NN
interactions. Therefore we had the hope to obtain a good description of finite
nuclei using this interaction model. We will demonstrate below that this hope
has been fulfilled to a good extent.

Because of the non-locality of $V_{lowk}$ nuclear structure calculations of finite
nuclei using this interaction are typically not performed in coordinate space but
employ a set of appropriate single-particle wave functions as a set of basis states,
typically those of a harmonic oscillator potential. The choice of the harmonic
oscillator (HO) basis seems to be rather plausible if one wants to describe the 
structure deeply bound, double magic nuclei like $^4He$, $^{16}O$, and  $^{40}
Ca$\cite{coraggio:2005,coraggio:2006}. This choice may be questionable if nuclei
close to the proton or neutron drip line are to be considered. The HO
basis may not be appropriate to describe the tail of the single-particle
wave functions for the weakly bound valence nucleons. The main reason, why this
oscillator basis is so popular, must be attributed to the fact that the HO basis
provides a simple transformation from relative coordinates in coordinate or
momentum space to a coordinate system located in a laboratory frame. Such a
transformation is required to allow for structure calculation of finite nuclei
using realistic interaction as defined e.g. by $V_{lowk}$.

The HO basis is certainly not appropriate to describe
the  quasi-nuclear structures in the pasta phase of neutron stars mentioned
above. For those studies a basis of plane wave (PW) states confined to a finite
box with a size large to the nuclear dimension may be more suitable and could be
considered as an appropriate basis for a Wigner-Seitz cell 
calculation\cite{Montani:2004,goegeleina:2008}. One of the main aims of this work
is to establish the techniques, which are needed to allow for nuclear structure
calculations of finite nuclear systems using realistic interactions in such a PW
basis.     

In this work we describe double magic nuclei and weakly bound neutron-rich
nuclei in the framework of Hartree-Fock (HF) plus BCS calculation using a
$V_{lowk}$ potential supplemented with a contact interaction.  We are going to
investigate nuclei up to $^{208}Pb$ and compare the results of calculations 
using HO and PW basis states.   

The plan of this paper is as follows. The procedure to renormalize the low-momentum
interaction and obtain the matrix elements of $V_{lowk}$ is reviewed in
Sec.~\ref{sec:Vlowk_mat_el}. In this section we also describe our interaction model
supplementing $V_{lowk}$ by an appropriate contact interaction. We also work out
how to solve the HF equation in a basis of harmonic oscillator and plane wave
states. This section is completed by a short description of the BCS procedure. The
results for infinite matter and finite nuclei are presented and discussed in
Sec.~\ref{sec:R&D}. Finally, Sec.~\ref{sec:SC} contains a summary and the
conclusions of our work.

\section{Model of the NN interaction}
\label{sec:Vlowk_mat_el}
The idea behind the  $V_{lowk}$ interaction model is to separate the low-momentum 
and high-momentum components of realistic interactions and restrict the nuclear
structure calculation to the low-momentum components. This means that one tries to
define a model space, which accounts for the low-momentum degrees of freedom and
renormalizes the effective Hamiltonian for this low-momentum regime to account for
the effects of the high-momentum parts, which are integrated out. A model space
technique to disentangle these parts, the unitary-model-operator approach (UMOA)
\cite{suzuki:1982}, can be used to calculate the matrix elements for $V_{lowk}$. 

First,
we define the projection operators $P$, which projects onto the low-momentum
subspace, and $Q$, which projects onto the complement of this subspace, as in all
well known model space techniques. Furthermore, these operators satisfy the usual
relations like $P+Q=1$, $P^2=P$, $Q^2=Q$, and $PQ=0=QP$. The idea of the
unitary-model-operator approach is now to define a unitary transformation $U$ so
that the transformed Hamiltonian does not couple $P$ and $Q$, which means
\begin{equation}
 QU^{-1}HUP = 0
\end{equation}
has to be fulfilled.
The effective interaction is defined in terms of this unitary transformation as
\begin{equation}
  V_{eff} = U^{-1}(h_0 + v_{12}) U - h_0,
\end{equation}
with $v_{12}$ representing the bare $NN$-interaction
and a starting Hamiltonian $h_0$ describing
the one-body part of the two-body system.
The unitary operator $U$ can be written as
\begin{equation}
  U = (1+ \omega -\omega^\dagger )(1+\omega \omega^\dagger + \omega^\dagger
\omega)^{-1/2}
\label{eq:umoa}
\end{equation}
with an operator $\omega$ fulfilling
$\omega = Q\omega P$ and $\omega^2 = \omega^{\dagger 2} = 0$.
In \cite{bozek:2006} the operator $\omega$  is  obtained
by first solving the two-body eigenvalue
equation
\begin{equation}
  (h_0 + v_{12} ) |\Phi_k \rangle = E_k | \Phi_k \rangle
\end{equation}
and afterwards defining the matrix elements of $\omega$
using the eigenstates $|\Phi_p \rangle$ having the largest overlap
with the $P$ space.
With the corresponding $U$ then the effective interaction $V_{eff}$ is calculated
as described in \cite{fuji:2004,bozek:2006}. In this way one obtains an effective
Hamiltonian $H_{eff} = h_0 + V_{eff}$, which contains this effective interaction
$V_{eff}$. The eigenvalues, which are obtained by diagonalizing this effective
Hamiltonian in the model-space ($P$-space), are identical to the diagonalization
of the original Hamiltonian $h_0+V$ in the complete space. 

This model-space scheme can now be applied to the effective two-nucleon problem by
considering for the basis states of the 2-nucleon system the states identified by
the relative momentum, its modulus and the corresponding partial waves. For a given
partial wave the states of the $P$-space are identified as those states with a
relative momentum smaller than a cutoff $\Lambda$. Applying the technique
described above leads to the effective interaction $V_{lowk}$. Solving now the
Schroedinger equation for $V_{lowk}$ within the model space, e.g. by solving the
Lipmann Schwinger equation for $NN$ scattering with the cutoff $\Lambda$, yields 
the same phase shift as solving the Lipmann Schwinger equation without cutoff for
the original interaction $v_{12}$. If the original interaction $v_{12}$ is
realistic in the sense, that it has been fitted to the experimental NN 
phase-shifts, these phase shifts, up to the cutoff $\Lambda$, will also be
reproduced by $V_{lowk}$.

Furthermore, the resulting $V_{lowk}$ is found to be essentially
model independent. This means that the  potential is independent of the
underlying realistic interaction $v_{12}$, if the cutoff $\Lambda$ is chosen around
$\Lambda$ = 2 fm$^{-1}$. Thus, one is able to extract a low-momentum potential
$V_{lowk}$, which in a model independent manner describes the low-momentum
component of realistic NN interactions in a more or less unique way. Note, 
that, by construction, this $V_{lowk}$ interaction is defined in terms of 
matrix elements in a basis of NN states labelled by relative momentum. 

The fact that the high-momentum or short-range components of realistic NN
interactions have been integrated out by means of the unitary transformation of
Eq.~(\ref{eq:umoa}), seems to lead to  problems in the description of the
saturation behavior of nuclear matter. Because of this lack of short-range
correlation effects, which are modified in the medium, the emergence of a
saturation point is prevented in calculations of symmetric nuclear
matter~\cite{bogner:2005,kuckei:2003}. In order to achieve saturation in symmetric
nuclear matter, 
three-body interaction terms or density-dependent two-nucleon interactions are
needed. Therefore, the effective interaction $V_{lowk}$ is supplemented by a simple
contact interaction, which we have chosen following the notation of the Skyrme
interaction to be of the form
\begin{equation}
\Delta\mathcal{V} = \Delta\mathcal{V}_0 + \Delta\mathcal{V}_3,
\label{eq:contact}
\end{equation}
with
\begin{equation}
\Delta\mathcal{V}_0 = \frac{1}{4}t_0[(2+x_0)\rho^2-(2x_0+1)(\rho_n^2+\rho_p^2)]
\end{equation}
and
\begin{equation}
\Delta\mathcal{V}_3 = \frac{1}{24}t_3 \rho^{\alpha}[(2+x_3)\rho^2-(2x_3+1)(\rho_n^2+\rho_p^2)],
\end{equation}
where $\rho_p$ and $\rho_n$  refer to the local densities for protons and neutrons
while the matter density is denoted as $\rho=\rho_p+\rho_n$. 
The parameters of the contact interaction are $t_0$, $x_0$, $t_3$, $x_3$ and 
$\alpha$.
As described below we have chosen a fixed value of $\alpha=0.5$ and $x_0=0.0$ and 
fitted $t_0$ and
$t_3$ and $x_3$ in such a way that HF calculations using $V_{lowk}$ plus the
contact term of Eq.~(\ref{eq:contact}) yield the empirical saturation point for
symmetric nuclear matter and reproduce the symmetry energy at saturation density.

%
%\section{Hartree-Fock Hamiltonian}
%\label{sec:HF_Hamil}
%
Within the HF approximation this interaction model leads to a single-particle 
Hamiltonian for protons and neutrons ($\nu=n,p$) of the form
\begin{equation}
H_{HF,\nu}=H_{kin}+H_{lowk,\nu}+\Delta H_{ct,\nu} + H_{Coul} \delta_{\nu p},
\label{eq:HFHamil}
\end{equation}
where $H_{kin}$ is the kinetic part and $\Delta H_{ct,\nu}$ originates from 
the contact interaction of Eq.~(\ref{eq:contact}). This part is given by
\begin{eqnarray}
\Delta H_{ct,\nu}=& t_0/2 [(2+x_0) \rho-(1+2 x_0) \rho_\nu]+t_3/24 [(2+x_3) 
(2+\alpha)
\rho^{1+\alpha} \nonumber\\ &- (2 x_3+1)  \{2 \rho^{\alpha} \rho_\nu + 
\alpha \rho^{\alpha-1} (\rho_n^2+\rho_p^2) \}].
\end{eqnarray}
Furthermore, for the charged particles, the protons,  one has an additional 
contribution to the hamiltonian, the Coulomb contribution. It is given by
\begin{equation}
H_{Coul}=U_{coul,dir}+U_{Coul,exch},
\end{equation}
where the direct term is
\begin{equation}
U_{coul,dir}=4 \pi e^2 \left\{ \begin{array}{l c c}
\int dr' r'^2 \rho_p(r')/r  & for    & r'\leq r \\
\int dr' r' \rho_p(r') & for &  r' \ge r
\end{array} \right.
\end{equation}
and the exchange term is
\begin{equation}
U_{coul,exch}=- e^2 \left(\frac{3}{\pi}\right)^{1/3} \rho^{1/3}_p.
\end{equation}
Note that $\Delta H_{ct,i}$ and $H_{Coul}$ are local. They are defined in terms of
the single-particle densities $\rho_i$ resulting from the eigenstates of 
$H_{HF,i}$, which implies that they have to be determined in a self-consistent way,
as usual for HF calculations. 
%This total Hamiltonian of Eq.~(\ref{eq:HFHamil}) has to be diagonalized in each iteration. From this diagonalization procedure the eigenfunctions are determined in terms of the basis functions of the used basis system, which will be described in the following section.
%The total energy can be calculation after the last iteration in the Hartree-Fock calculation
%
%\begin{equation}
%E=E_{kin}+E_{vlowk}+E_{ct}+E_{Coul}+E_{pair}-E_{cm}
%\end{equation}
%
%\begin{equation}
%E=\left( \sum_{\beta} (\epsilon_{kin,\beta}+\epsilon_{sp,\beta}) \omega_{\beta} \right) + E_{ct,rearr} + E_{pair} - E_{cm}
%\end{equation}
%

%
%\section{The basis systems}
%\label{sec:basis}

As it has been mentioned above, the effective interaction $V_{lowk}$ is non-local
and defined in terms of matrix elements. This implies that the HF calculation has
to be performed in a Hilbert space using an appropriate basis $\vert \alpha >,
\vert \beta >, ...$. The HF Hamiltonian is then expressed in terms of the matrix
elements between these basis states $< \alpha \vert H_{HF} \vert \beta >$ and
the HF single-particle states $\vert \Psi_n >$ are defined in terms of the
expansion coefficients in this basis
\begin{equation}
\vert \Psi_n > = \sum_{\alpha} \vert \alpha > < \alpha \vert\Psi_n > = 
\sum_{\alpha}
c_{n\alpha} \vert \alpha >\,. 
\end{equation}
The part of the HF Hamiltonian originating from $V_{lowk}$ can be expressed in
terms of two-body matrix elements by
\begin{equation}
< \alpha \vert H_{lowk} \vert \beta > = \sum_{\gamma,\delta} < \alpha \gamma\vert 
V_{lowk} \vert \beta \delta> \rho_{\gamma\delta}\,,\label{eq:defhlowk}
\end{equation}
with $\rho_{\gamma\delta}$ representing the single-particle density matrix.

If we restrict the HF variational procedure to a spherical description of nuclei,
as we will do in the following, the HF Hamiltonian is diagonal in the angular
momentum quantum numbers $j,l,m$ and the expansion is simplified to 
\begin{equation}
<\vec r \vert \Psi_n > = 
\Psi_{nljm}(\vec r) = \sum_{i=1}^{\infty} c_{nilj}
\Phi_{iljm}(\vec r) \approx \sum_{i=1}^Nc_{nilj}
\Phi_{iljm}(\vec r)\,,
\label{eq:hfbasis}
\end{equation}
where $c_{nilj}$ are expansion coefficients and $\Phi_{iljm}(\vec r) = <\vec r
\vert \alpha >$ are the wave functions of our orthogonal basis. The number of basis
states $N$ has to be chosen to guarantee that the results are not affected by this
limitation. Note, that using this coordinate space representation of the basis 
states allows the calculation of the matrix elements of $< \alpha \vert H_{HF} 
\vert \beta >$ for local terms in $H_{HF}$ in a straight forward way.

In this paper calculations are done using two different orthogonal
basis systems. These basis systems are respectively that of a free particle and 
that of a spherical harmonic oscillator. In these cases the wave functions of our
orthogonal basis can be separated in a radial part and an angular part,
\begin{equation}
\Phi_{iljm}(\vec r) = \langle \vec r \vert iljm\rangle = R_{il}(r)
{\cal Y}_{ljm} (\vartheta,\varphi)\,.\label{eq:basis}
\end{equation}
In both basis systems, ${\cal Y}_{ljm}$ represent the spherical harmonics including
the spin degrees of freedom by coupling the orbital angular momentum $l$ with
the spin to a single-particle angular momentum $j$.

The radial part $R_{il}$ is different for the two basis systems.
The first basis system considered is that of a particle moving free in a spherical
cavity with a radius $R$, the plane wave (PW)
basis~\cite{Montani:2004}. The radial wave functions $R_{il}$ are then given by the
spherical Bessel functions,
\begin{equation}
R_{il}(r) = N_{il} j_l(k_{il}r),
\label{eq:PW_Radial}
\end{equation}
for the discrete momenta $k_{il}$, which fulfill
\begin{equation}
R_{il}(R)= N_{il}j_l(k_{il}R) = 0 \ => \ j_l(k_{il}R) = 0\, .\label{eq:PW_condition}
\end{equation}
The normalization constant in Eq.~(\ref{eq:PW_Radial}) is given by
\begin{equation}
N_{il}=\left\{\begin{array}{ll}
i\pi \frac{\sqrt{2}}{\sqrt{R^3}}&\mbox{for}\,l=0\,,\\
\frac{1}{j_{l-1}(k_{il}R)} \frac{\sqrt{2}}{\sqrt{R^3}} &
\mbox{for}\,l>0\,.\end{array}\right.
\end{equation}
It ensures that the basis functions are orthogonal and normalized within the box,
\begin{equation}
\int_0^R d^3r \Phi^*_{iljm}(\vec r)\Phi_{i'l'j'm'}(\vec r) =
\delta_{ii'}\delta_{ll'}\delta_{jj'}\delta_{mm'}\,.
\end{equation}

Next, the basis system of the harmonic oscillator (HO) is considered.
The radial part of the basis functions takes the form~\cite{moshinsky:1959}
\begin{equation}
R_{il}(r)=\sqrt{\frac{2(i!)}{a^3 \ \ \Gamma(i+l+3/2)}} \ (r/a)^l \ \exp(-0.5 \
r^2/a^2) \ L_i^{l+1/2}(r^2/a^2), \label{eq:HO_Radial}
\end{equation}
where $L_i^{l+1/2}$ is a Laguerre polynomial, $\Gamma$ is a gamma function, and
$a=\sqrt{\hbar/(m\omega)}$ is the oscillator length. These basis functions
$\Phi_{iljm}$ are orthogonal and normalized within a box with an infinite radius.
However, in practice one can use a radius $R>>a$ due to the exponential factor
$\exp(-0.5 \ r^2/a^2)$ in the radial basis function of  Eq.~(\ref{eq:HO_Radial}).

The matrix elements of $V_{lowk}$ or other realistic NN interactions,
which are required in Eq.(\ref{eq:defhlowk}), are easily calculated within the HO
basis: In a first step one can calculate the oscillator matrix elements using the
momentum representation of the relative basis. The second step, transforming these
matrix elements from the relative system to the laboratory system is also straight
forward using the well known transformation brackets of the Talmi-Moshinsky
transformation\cite{moshinsky:1959,talmi:1952,talmi:1993}.

While the corresponding first step, the calculation of matrix elements of
$V_{lowk}$ in the relative basis is trivial in the case of PW states, the
transformation from relative to laboratory coordinates is much less convenient in
the PW than in the HO basis. The evaluation and use of these vector brackets has
been described e.g. in \cite{Kung:1979,bonatsos:1989} and will be used here. 

%
%\section{BCS Pairing.}
%\label{sec:BCS}
%
The Hartree-Fock approximation discussed so far is a good approximation for double
closed shell nuclei. In order to consider nuclei with open shells as well, we are
going to use a rather phenomenological treatment of pairing correlations.
The pairing is described by a BCS approach with a smooth cutoff. For the BCS
potential $V(r)$ we have taken a density dependent effective interaction of 
zero range as introduced by Bertsch and Esbensen\cite{DP:Bertsch91},
\begin{equation}
V(r)=V_0 \ [1.0-\eta \ (\rho(r)/\rho_0)^{\beta}],
\label{eq:BCSpot}
\end{equation}
where $\rho_0=0.16$ fm$^{-3}$ is the saturation density. Furthermore, the values
$V_0=481.0$, $\eta=1.0$, and $\beta=1.0$ are taken for the parameters $V_{0},
\eta,$ and $\beta$. These parameters have been determined by Garrido {\it et
al.}\cite{PP:Garrido99}. 

The anomalous density has the form of
\begin{equation}
\chi(r)=\sum_{nljm} (2j+1) \frac{w_{nljm} u_{nljm} v_{nljm}}{2}  \psi_{nljm}(r)^2,
\label{eq:anoden}
\end{equation}
where $u_{nljm}$ and $v_{nljm}$ are variational parameters solved from the BCS
equations,
\begin{equation}
u_{nljm}^2=\frac{1}{2}\left[1+\frac{\epsilon_{nljm}-\epsilon_F}{E_{nljm}}\right];
\quad v_{nljm}^2=\frac{1}{2}\left[1-\frac{\epsilon_{nljm}-
\epsilon_F}{E_{nljm}}\right]
\end{equation}
with $E_{nljm}=\sqrt{(\epsilon_{nljm}-\epsilon_F)^2+\Delta_{nljm}^2}$ the energy of
the quasi-particle state and $\epsilon_{nljm}$ the single-particle energy
determined in the HF equations.  Furthermore, $\Psi_{nljm}(\vec r)$ is the HF
orbital wave function of Eq.~\ref{eq:hfbasis} and $w_{nljm}$ is the smooth cutoff,
\begin{equation}
w_{nljm}=[1.0+\exp((\epsilon_{nljm}-E_F-E_{cut})/\Delta\epsilon)]^{-1},
\end{equation}
with $E_{cut}=5$ MeV and $\Delta\epsilon=E_{cut}/10$.  Therefore, the local gap
function can be written as
\begin{equation}
\Delta(r)=V(r)\chi(r),
\end{equation}
where $V(r)$ is the BCS potential of  Eq.~\ref{eq:BCSpot} and $\chi(r)$ is the
anomalous density in Eq.~(\ref{eq:anoden}).

\section{Results and Discussion}
\label{sec:R&D}
\subsection{Nuclear Matter}
\label{sec:nucl_mat}

All the results which we are going to discuss in this chapter employ a
low-momentum interaction $V_{lowk}$, which is based on the proton-neutron part 
of CD Bonn potential\cite{mach1}. Using a cutoff parameter $\Lambda$ = 2
fm$^{-1}$, these results do not significantly depend on this choice of the 
underlying realistic interaction. 

\begin{figure}
\begin{center}
\includegraphics[width=0.6\textwidth]{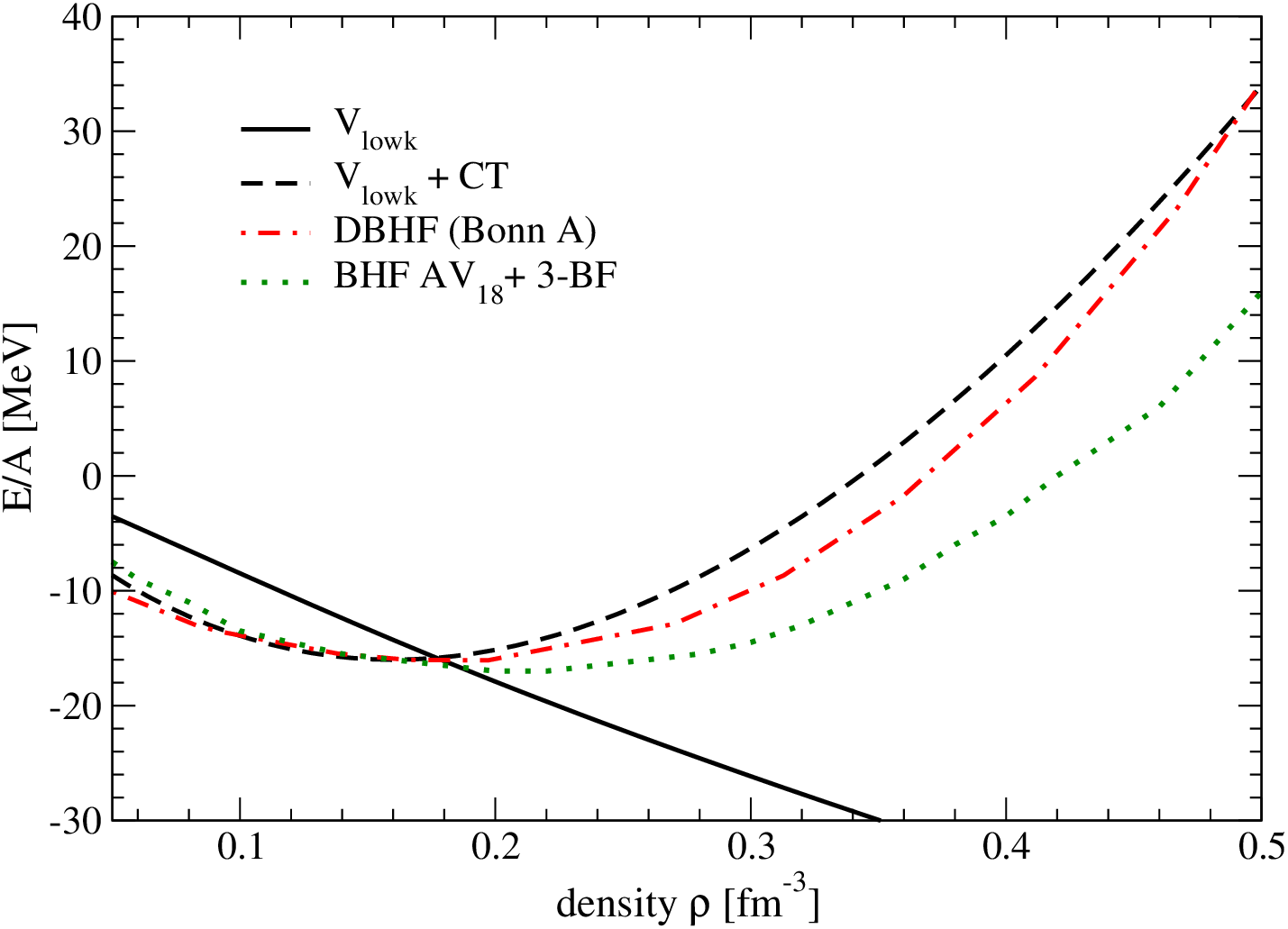}
\end{center}
\caption{(Color online) Binding energy per nucleon of symmetric nuclear matter of
a Hartree-Fock calculation using $V_{lowk}$ (solid line) and of a calculation
supplemented with a contact interaction (dashed line). Furthermore, comparison is
made with a Dirac Brueckner Hartree Fock (DBHF)
calculation~\cite{vandalen:2004b,vandalen:2007} (dashed-dotted line) and a
Brueckner Hartree Fock calculation using the potential Argonne V18 plus a
3-nucleon force (BHF + 3BF)~\cite{lejeune:2000} (dotted line).}
\label{fig:SymMEb}
\end{figure}
First, let us turn to the binding energy of symmetric nuclear matter, which are
displayed in Fig.~\ref{fig:SymMEb}. The HF calculation using $V_{lowk}$ does not
lead to a minimum in the energy versus density plot (see solid line in
Fig.~\ref{fig:SymMEb}), as we have already mentioned
before~\cite{bogner:2005,kuckei:2003}. The absence of saturation is one of the
major problems of calculations for nuclear matter using $V_{lowk}$. This problem
cannot be cured by the inclusion of correlations beyond the HF 
approximation\cite{bozek:2006}. Using a conventional model for a realistic
interaction, like e.g. the CD Bonn interaction or the Argonne V18, which are not
reduced to its low-momentum components, one obtains a saturation point in
calculations which account for correlations beyond the mean-field approximation.
Therefore one may argue that the $V_{lowk}$ approach cannot reproduce the
saturation of nuclear matter as it misses the quenching of short-range
correlations in the nuclear medium, which is included in sophisticated
calculations using one of the conventional models for a realistic NN interaction.

Note, however, that calculations like Brueckner Hartree Fock (BHF) for these
interactions lead to a saturation point, but are not able to reproduce the
empirical data. Depending on the amount of tensor and short-range correlations
produced by these interactions either the predicted density is too large or the
calculated binding energy is too weak\cite{Muether00,coester:1970}. In order to
obtain the experimental data for the saturation point one has to include a
three-body force\cite{lejeune:2000} or include relativistic effects e.g. within
a Dirac Brueckner Hartree Fock (DBHF) 
approach\cite{brock:1984,muemach:1990,vandalen:2004b,vandalen:2007}. The
relativistic effects due to the change of the nucleon Dirac spinors in the
medium must be considered within the framework of non-relativistic calculations
by means of a density-dependent two-nucleon or an effective three-nucleon
interaction.

In Fig.~\ref{fig:SymMEb}, our calculation for the binding energy is compared to
the ones of a DBHF calculation from Ref.~\cite{vandalen:2004b,vandalen:2007} and
the non relativistic BHF calculation from Ref.~\cite{lejeune:2000}.  The
non relativistic BHF calculation is based on the Argonne $V_{18}$~\cite{arg1} and
includes a three-body force deduced from the meson-exchange current
approach~\cite{lejeune:2000}.

Using the $V_{lowk}$ approach, the absence of saturation can be cured by adding
three-body forces~\cite{bogner:2005}. Therefore, we also  supplemented the
$V_{lowk}$ potential with the contact interaction of Eq.~(\ref{eq:contact}) to
obtain the experimental saturation behavior with a binding energy per nucleon of
$E/A=-16.0$ MeV at a saturation density $\rho_0$=0.16 fm$^{-3}$. It is worth to
notice that only the isoscalar part of the contact interaction can influence the
binding energy in symmetric nuclear matter. 

\begin{figure}
\begin{center}
\includegraphics[width=0.6\textwidth]{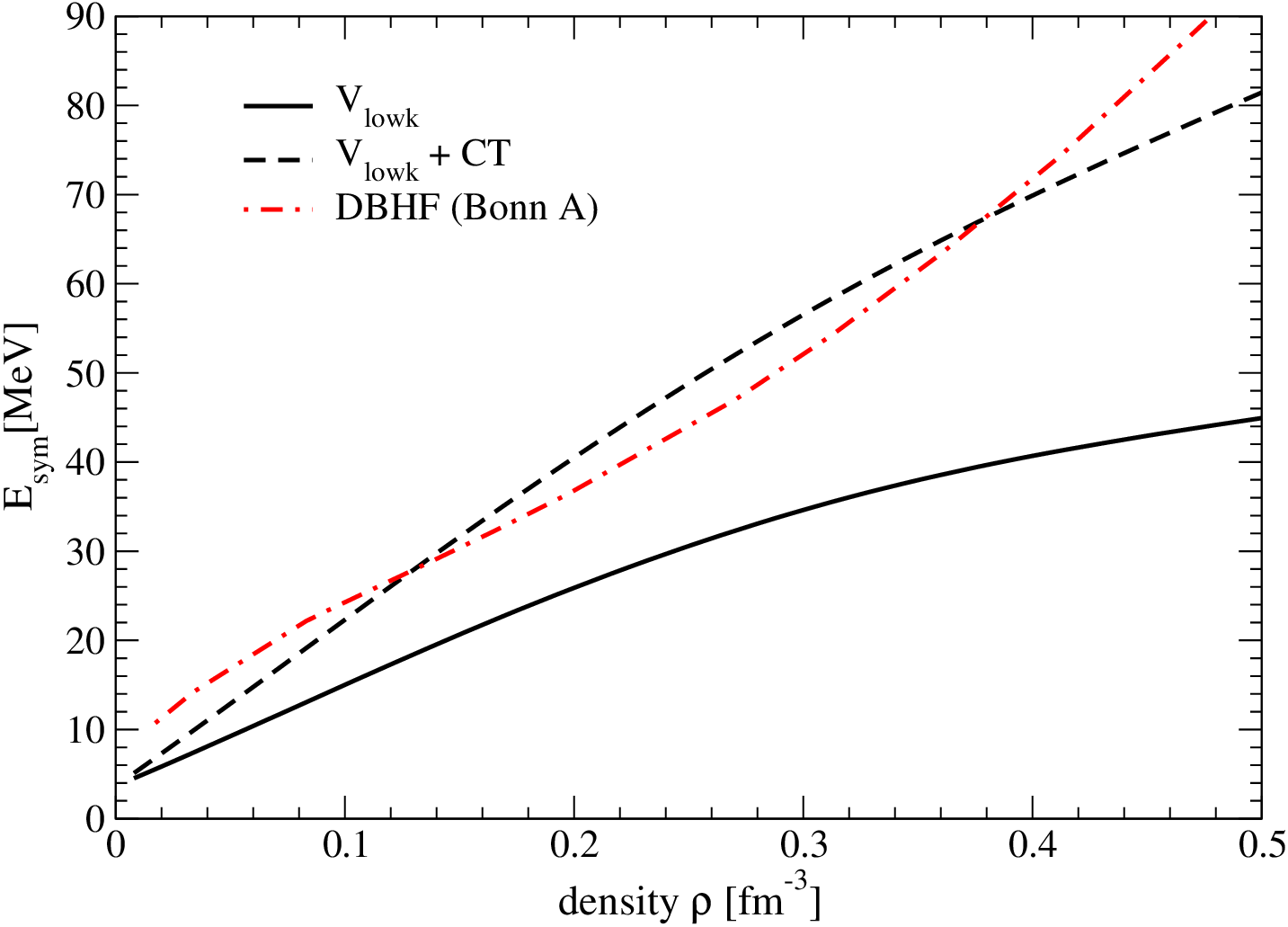}
\end{center}
\caption{(Color online) The symmetry energy as a function of the density. The
symmetry energy of the Hartree-Fock calculations using both $V_{lowk}$ (solid
line) and $V_{lowk}$ plus a contact interaction (dashed line) is plotted. For
comparison the results of the DBHF approach~\cite{vandalen:2007}
 are plotted as well.}
\label{fig:SymEVlowk}
\end{figure}

In order to fix the isovector part of the contact interaction as well, we also
inspect the symmetry energy of infinite matter as a function of density in 
Fig.~\ref{fig:SymEVlowk}. The symmetry energy of 21 MeV at saturation density
obtained from the $V_{lowk}$ calculation  is only about two third of the
experimental value. Furthermore, the value of the symmetry energy is significantly
below the one predicted by DBHF also at higher densities. This means that the
symmetry energy, respectively the isospin dependence of the EoS, of $V_{lowk}$
without contact interaction is in principal too soft.

So we
have added a contact interaction term as defined in Eq.~(\ref{eq:contact})
choosing a value for $\alpha$ =0.5 and $x_0=0.0$. The three remaining parameters
have then been fitted to reproduce the empirical saturation point of nuclear
matter and the symmetry energy at saturation density. 
The results for these fitting parameters are listed in
Table~\ref{table:CT_Param} and the corresponding results for the binding energy of
symmetric nuclear matter and the symmetry energy as a function of density are
displayed in Figs.~\ref{fig:SymMEb} and  \ref{fig:SymEVlowk} as well.

% fit parameters
\begin{table}
\begin{center}
\begin{tabular}{|c|c|c|c|}
\hline
\ \ $t_0$ \ [MeV fm$^3$] \ \ & \ \ \ \ \ \ \ \ $x_0$  \ \ \ \ \ \ \ \ & \ \ $t_3$
\ [MeV fm$^{3+3\alpha}$] \ \ & \ \ \ \ \ \ \ \
$x_3$ \ \ \ \ \ \ \ \ \\
\hline
584.1 & 0.0 & 8330.7 & -0.5   \\
\hline
\end{tabular}
\end{center}
\caption{\label{table:CT_Param}
Parameters $t_0$, $x_0$, $t_3$ and $x_3$ defining the contact interaction of
Eq.~(\protect{\ref{eq:contact}})}
\end{table}

The first observation that becomes evident from Fig.~\ref{fig:SymMEb} is that the
$V_{lowk}$ plus contact interaction ($V_{lowk}$+CT) and DBHF calculations yield
results which are rather similar, while the result of the BHF approach predicts
lower energies at high densities.  The compressibility modulus of $V_{lowk}$+CT at
saturation density is $K=258$ MeV, while those of the other microscopic
calculations range from 200 MeV to 265 MeV. This means all the presented equations
of state (EoSs) in Fig.~\ref{fig:SymMEb} can be characterized as rather soft, at
least at densities up to about three times saturation density. This prediction of
a soft equation of state (EoS) is also supported from observables extracted from
heavy ion reactions. For example, heavy ion data for transverse
flow~\cite{stoicea:2004} or from kaon production~\cite{sturm:2001} support the
picture of a soft EoS in symmetric nuclear matter.

In Fig. \ref{fig:SymEVlowk} we can see that the addition of the contact
interaction enables us to bring the value of the symmetry energy close to the
experimental value. The symmetry energy at saturation density is now 33 MeV.
Moreover it is comparable to that of the microscopic DBHF approach up to densities
of three times saturation density. Therefore the isospin dependence of the EoS for
$V_{lowk}$+CT calculation is rather stiff.

\subsection{Finite Nuclei}
\label{sec:fin_nucl}

\begin{figure}
\begin{center}
\includegraphics[width=0.5\textwidth]{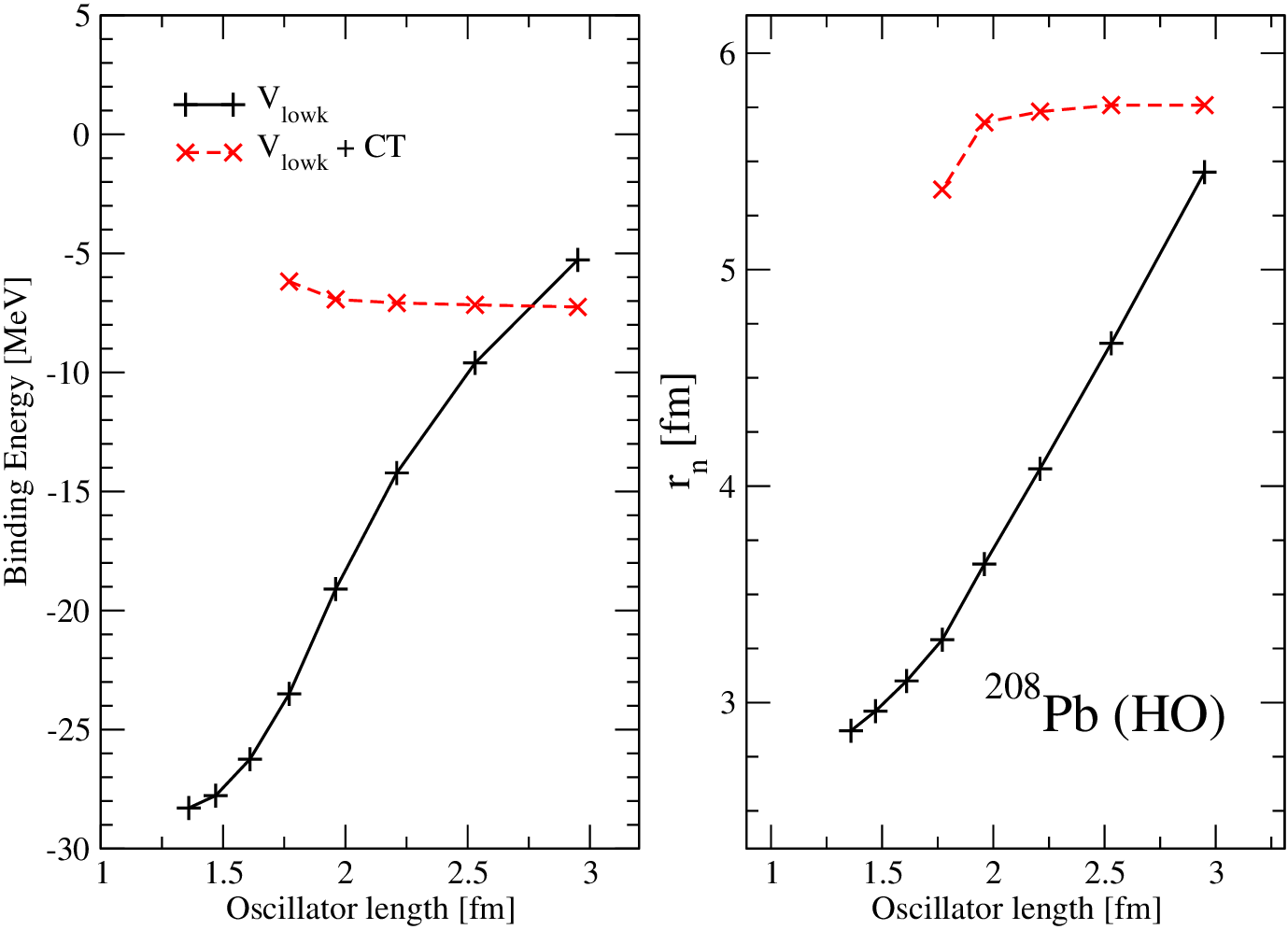}
\end{center}
\caption{(Color online) Results of Hartree Fock calculations for the energy per
nucleon (left panel) and the radius of the neutron distribution (right panel) of
$^{208}Pb$. The calculations use either the bare $V_{lowk}$ or the $V_{lowk}$ plus
contact (CT) interaction. The single particle wave functions are expanded in an
oscillator basis. Results are presented as a function of the oscillator length
parameter $a$.\label{fig:208stabi}} 
\end{figure}

In the first part of this section we are going to consider the bulk properties 
of nuclei with closed shells for protons and neutrons. The typical example for a
heavy nucleus is the example of $^{208}Pb$. Results for HF calculations using the
$V_{lowk}$ interaction, expanding the single-particle wave functions in a basis of
harmonic oscillator (HO) states are presented Fig.~\ref{fig:208stabi}. The
left panel of this figure displays the calculated binding energy as a function of
the oscillator length $a$ (see Eq.(\ref{eq:HO_Radial})). The expansion of the
single-particle wave functions has been restricted to oscillator states with $N
\leq 8$. One finds that the resulting energy is rather sensitive to the choice of
the oscillator parameter and leads to very attractive values at small values of
$a$. This gain in binding energy is accompanied by a decrease of the calculated
radius of the neutron distribution as can be seen from the right panel  in this
figure. The nucleus is collapsing to a tightly bound system with a large density.
So these results reflect the properties of infinite matter if the bare $V_{lowk}$
interaction is employed.

\begin{figure}
\begin{center}
\includegraphics[width=0.3\textwidth]{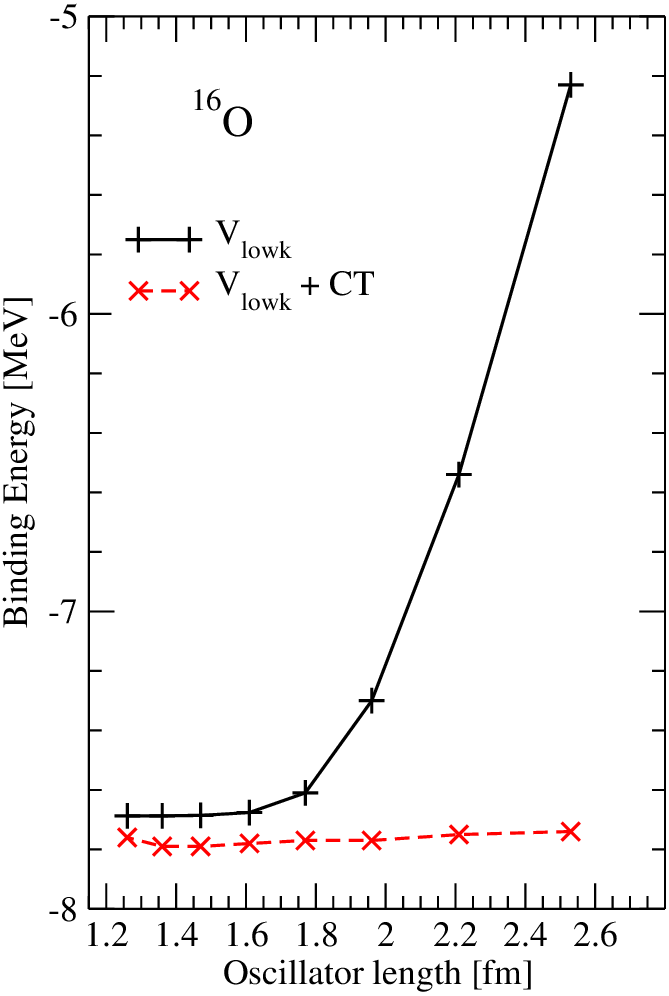}
\end{center}
\caption{(Color online) Results of Hartree Fock calculations for the energy per
nucleon of
$^{16}O$. The calculations use either the bare $V_{lowk}$ or the $V_{lowk}$ plus
contact (CT) interaction. The single particle wave functions are expanded in an
oscillator basis. Results are presented as a function of the oscillator length
parameter $a$.\label{fig:16stabi}} 
\end{figure}

The situation is rather different if we consider light nuclei. As an example we
focus our attention on $^{16}O$ and present in Fig.~\ref{fig:16stabi} the results
for the energy per nucleon using various values for the oscillator parameter $a$.
Using $V_{lowk}$ one obtains results which are rather insensitive on the choice of
$a$ over a large range of values. Only a basis with $a$ larger than 2 fm seems not
to be adequate for such calculation and leads to smaller values for the binding
energy per nucleon. Also it is worth noting that the calculated energies around
-7.7 MeV per nucleon are rather close to -7.98 MeV observed in experiment. This
rather good description of light nuclei using $V_{lowk}$ in an appropriate HO
basis has been observed before by Coraggio {\it et
al.}\cite{coraggio:2003,coraggio:2006}.  

How are these results for finite nuclei modified by the contact interaction, which
we have added to describe the properties of infinite matter? This is visualized
in Figs.~\ref{fig:208stabi} and \ref{fig:16stabi} as well. In the case of
$^{208}Pb$ the contact interaction provides a stabilization of the results leading
to values for the energy per nucleon and radius of the nucleon distribution, which
are in good agreement with the experimental data. 

The same is true as well for the case of $^{16}O$. The results obtained for the
energy per nucleon in HF calculation is very insensitive to the oscillator
parameter considered and the calculated energy is rather close to the experimental
value.

\begin{figure}
\begin{center}
\includegraphics[width=0.6\textwidth]{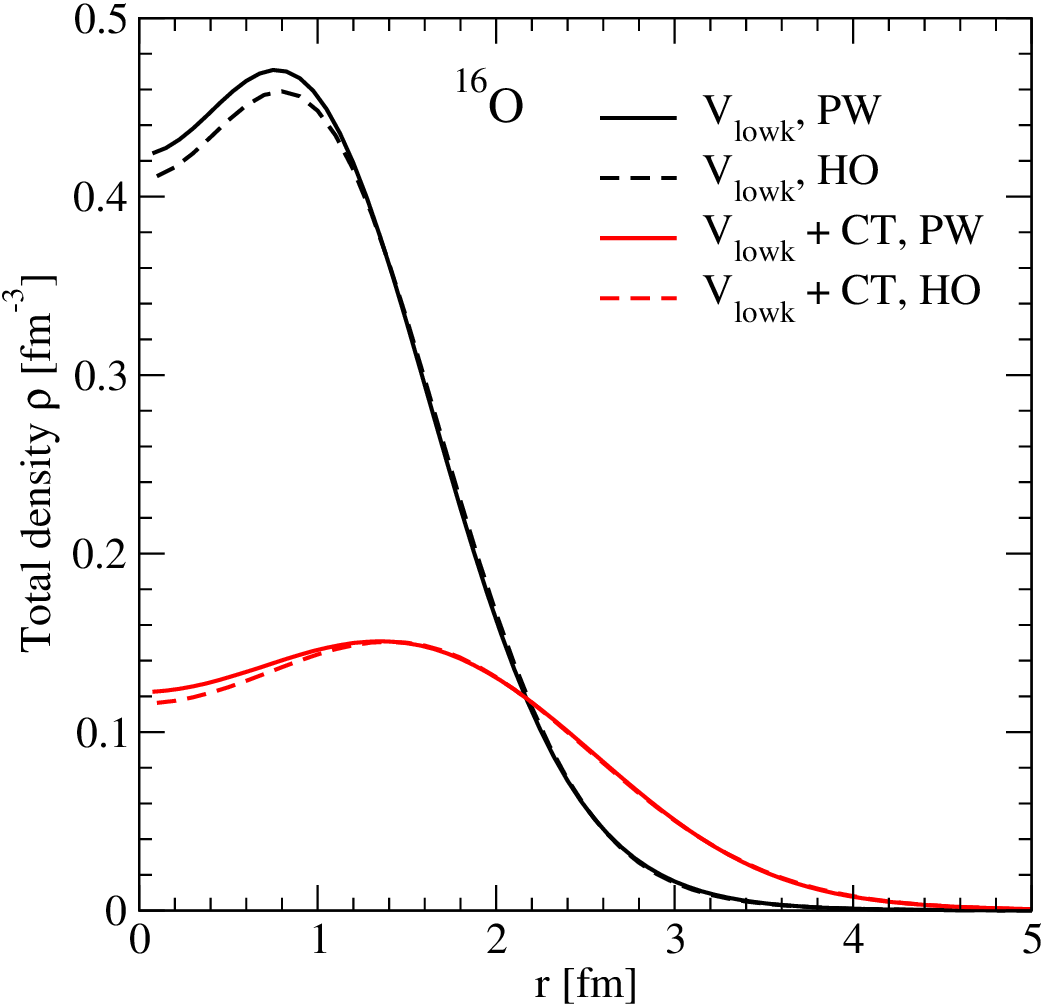}
\end{center}
\caption{(Color online) Comparison of total density distributions calculated
for  $^{16}O$ using plane wave basis in a spherical box of radius $R=15$ fm (PW,
solid lines) and an appropriate harmonic oscillator (HO, dashed lines) basis.
Results of HF  $V_{lowk}$ calculation without contact term (dark lines) and with
contact  term (light lines) are shown.}
\label{fig:densityO16}
\end{figure}

The calculated energies for $^{16}O$ with and without the contact interaction are
rather close to each other (see Fig.~\ref{fig:16stabi}), the resulting wave
functions, however, are quite different. This can be seen from 
Fig.~\ref{fig:densityO16} showing the calculated nuclear density profiles. 
The bare $V_{lowk}$ interaction yields densities in the center of the nucleus, 
which are close to three times the saturation density of nuclear matter.
Consequently, the calculated radii are very much below the empirical result. This
is in line with the HF calculations of \cite{coraggio:2003}. This also explains
why oscillator parameters $a$ larger than 2 fm, are not suitable to describe such
a solution. The density profile derived from the $V_{lowk}$ + CT interaction model
yields densities below the saturation density and consequently results for the
radius of the charge distribution, which are in good agreement with experiment
(see discussion below). 

Up to this point we have restricted the discussion to HF calculations performed
within an oscillator basis. How do these results compare with the ones derived
within the PW basis. Before we proceed to this comparison it should be mentioned
that the calculations within the HO basis are much easier to perform than those
within the PW basis. Not only that the techniques to calculate the two-body
matrix elements (in particular the transformation from relative to laboratory
coordinates) is more tediously, also the number of basis states which have to be
considered is typically larger in the PW basis as compared to the HO basis. In
order to obtain results which are stable from the numerical point of view, we
have typically considered a reference box with a radius $R$ which is about 5
times as large as the radius of the nucleus considered. In the case of $^{16}O$
we have typically considered boxes with radii $R$ = 15 fm, while a calculation
of $^{208}Pb$ seems to require a box as large as $R$ = 30 fm. This seems to be
necessary in order to obtain a set of radial momenta ($k_{il}$ in
Eq.(\ref{eq:PW_Radial})), which has a sufficient resolution. For small values of
the reference box we also run into the problem that one is considering
interactions between neighbored Wigner Seitz cells. For such large boxes,
however, we need a large number of such momenta to cover the relevant range of
momenta. This number is typically of the order 15 or larger. The number of
different oscillator states, which are required to obtain stable results is
significantly smaller.

Comparing the results displayed by dashed and dotted lines in 
Fig.~\ref{fig:densityO16} one find that the results for the density
distributions, which are obtained on the PW basis (solid lines) are very close
to the ones derived in the HF approximation using the HO basis. This is
particularly true for the calculations including the contact interaction term.
Note that the HF calculation using the PW basis leads to an energy per
nucleon of -7.91 MeV, which is slightly more attractive than the corresponding
number of -7.77 MeV obtained in the HO basis. This indicates that the larger
number of variational parameters, which are optimized in the PW basis (see
discussion above) leads to deeper minimum than the HF variational procedure in
the HO basis. Both results are rather close to each other and to the
experimental value (see results listed in table~\ref{table:Eb}).

\begin{table}
\begin{tabular}{|c|c c c|c c|c c |}
\hline
Nucleus    & &E/A [MeV]& & $r_{rms,p}$ & [fm] & $r_{rms,n}$ &[fm] \\
           & HO & PW & exp.& HO & PW &HO & PW \\
\hline
$^{16}O$   &  -7.77 &  -7.91 &  -7.98    & 2.68  & 2.68  &  2.66  &  2.66  \\
$^{40}Ca$  &  -8.33 &  -8.57 &  -8.55    & 3.41  &  3.41  &  3.37 &  3.37  \\
$^{48}Ca$  &  -8.13 &  -8.42 &  -8.67   & 3.46  &  3.45  &  3.68  &  3.68   \\
$^{60}Ca$  &  -7.42 &  -7.75 &    -    & 3.60  &  3.59   &  4.11  &  4.15  \\
$^{208}Pb$ &  -7.25 &  -7.76 & -7.87  & 5.48 & 5.45 & 5.76& 5.70\\
    \hline
    \end{tabular}
\caption{Results for the binding energies per nucleon and rms radii for 
different nuclei using the $V_{lowk}$ plus contact term interaction model using
the harmonic oscillator (HO) or plane wave (PW) basis. The calculated energies
have been corrected by subtracting the spurious energy of the center of mass
motion.
The experimental values are taken from Ref.~\cite{hofmann:2001}.}
\label{table:Eb}
\end{table}

This table exhibits very similar findings for the HF calculations of other
nuclei with closed shells for protons and neutrons including examples from light
($^{16}O$) to heavy nuclei ($^{208}Pb$). It seems that our fitting procedure for
the contact term in infinite matter yields very good results for the energies
and radii of finite nuclei. 

Also the single-particle energies are very similar using the HO or PW basis.
This can be seen from table \ref{table:O16}, which lists the single-particle
energies for the example of $^{16}O$ for the occupied 0s and 0p shells as well 
as for the 1s0d shell. The single-particle energies evaluated in the PW basis
are slightly more attractive than those derived within the HO approximation. The
difference is smaller of the tightly bound 0s states than for the more weakly
bound states in the 0p and 1s0d shells. 

\begin{table}
\begin{tabular}{|c c c c c c c c|}
\hline
    & & & & $^{16}O$ & & & \\
    & &Neutron& & & & Proton &  \\
Orbital  & HO & PW & exp. & & HO & PW & exp. \\
\hline
$s_{1/2}$   & -37.024 & -37.162  &  -47&& -33.478 & -33.601  &  -44 $\pm$7 \\
$p_{3/2}$  & -19.730 &  -20.006&  -21.839 && -16.358 & -16.632 &  -18.451  \\
$p_{1/2}$  & -16.217 & -16.484 &  -15.663 && -12.888 & -13.155 &  -12.127 \\
$d_{5/2}$  & -3.509 & -3.739 &    -4.144 && -0.444 & -0.690 &  -0.601   \\
$1s_{1/2}$   & -1.288 & -1.566 &  -3.273&& 1.436 &   0.839 &  -0.106  \\
$d_{3/2}$  & 0.720 & 0.339 &    0.941 && 3.562 &  1.886   &  4.399       \\
    \hline
    \end{tabular}
\caption{Single particle energies for the orbital levels of $^{16}O$ nucleus
derived from HF calculations using the $V_{lowk}$ plus contact interaction.
Results are displayed using the harmonic oscillator (HO) and plane wave (PW)
basis.
The experimental values are taken from Ref.~\cite{coraggio:2003}.}
\label{table:O16}
\end{table}

\begin{figure}
\begin{center}
\includegraphics[width=0.6\textwidth]{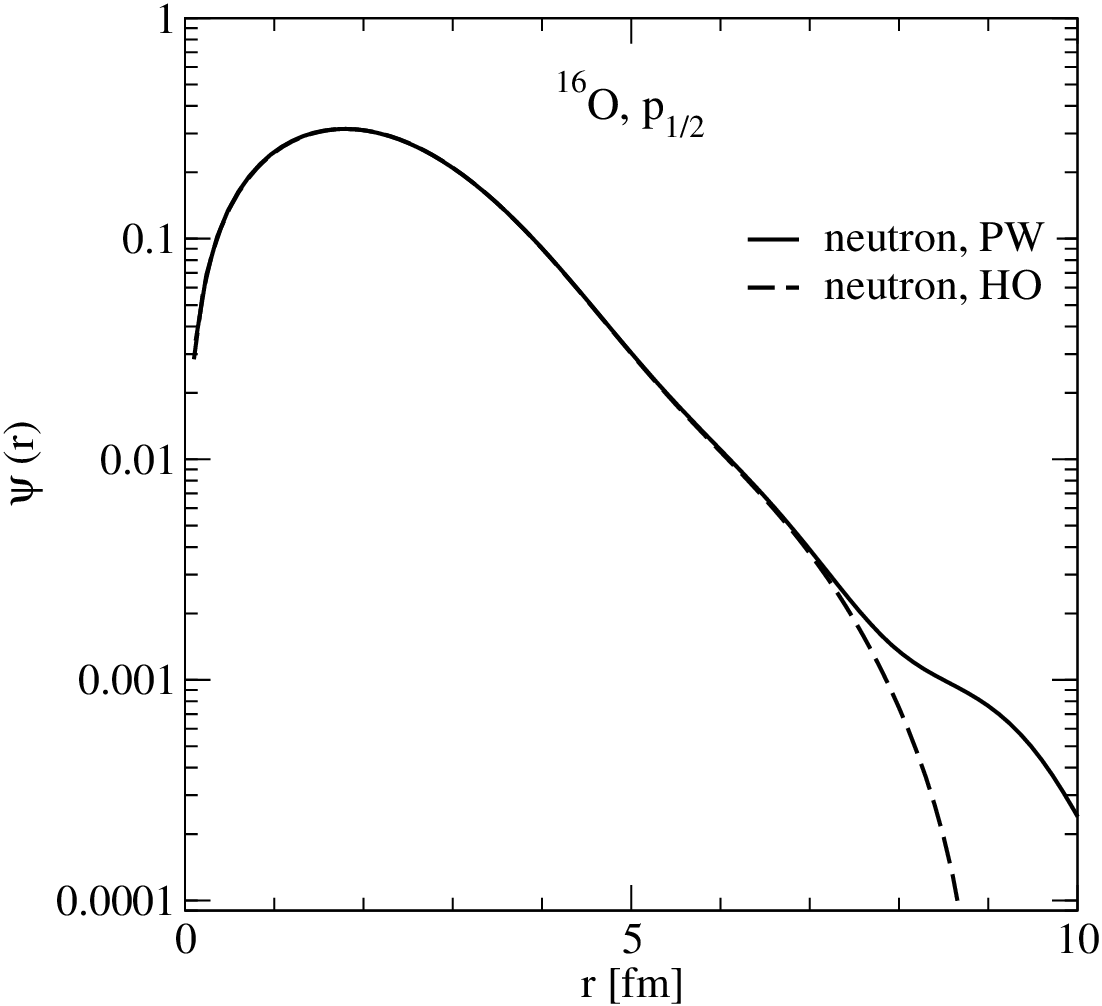}
\end{center}
\caption{The neutron  wave functions for the state $p_{1/2}$ states of 
$^{16}O$ calculated in the HF approximation using two different basis systems: 
the plane wave basis (solid lines) and the harmonic oscillator basis 
(dashed lines).}
\label{fig:O16R1p1}
\end{figure}

The shapes of corresponding single-particle wave function displayed in
Fig.~\ref{fig:O16R1p1} may exhibit the reason for this behavior. The wave
functions resulting from the HO and PW cannot be distinguished in this
logarithmic plot for radii up to 7 fm. For larger distance from the center,
however, the oscillator model yields a faster
decrease reflecting the Gaussian asymptotic, which is typical for the HO model.    

\begin{figure}
\begin{center}
\includegraphics[width=0.6\textwidth]{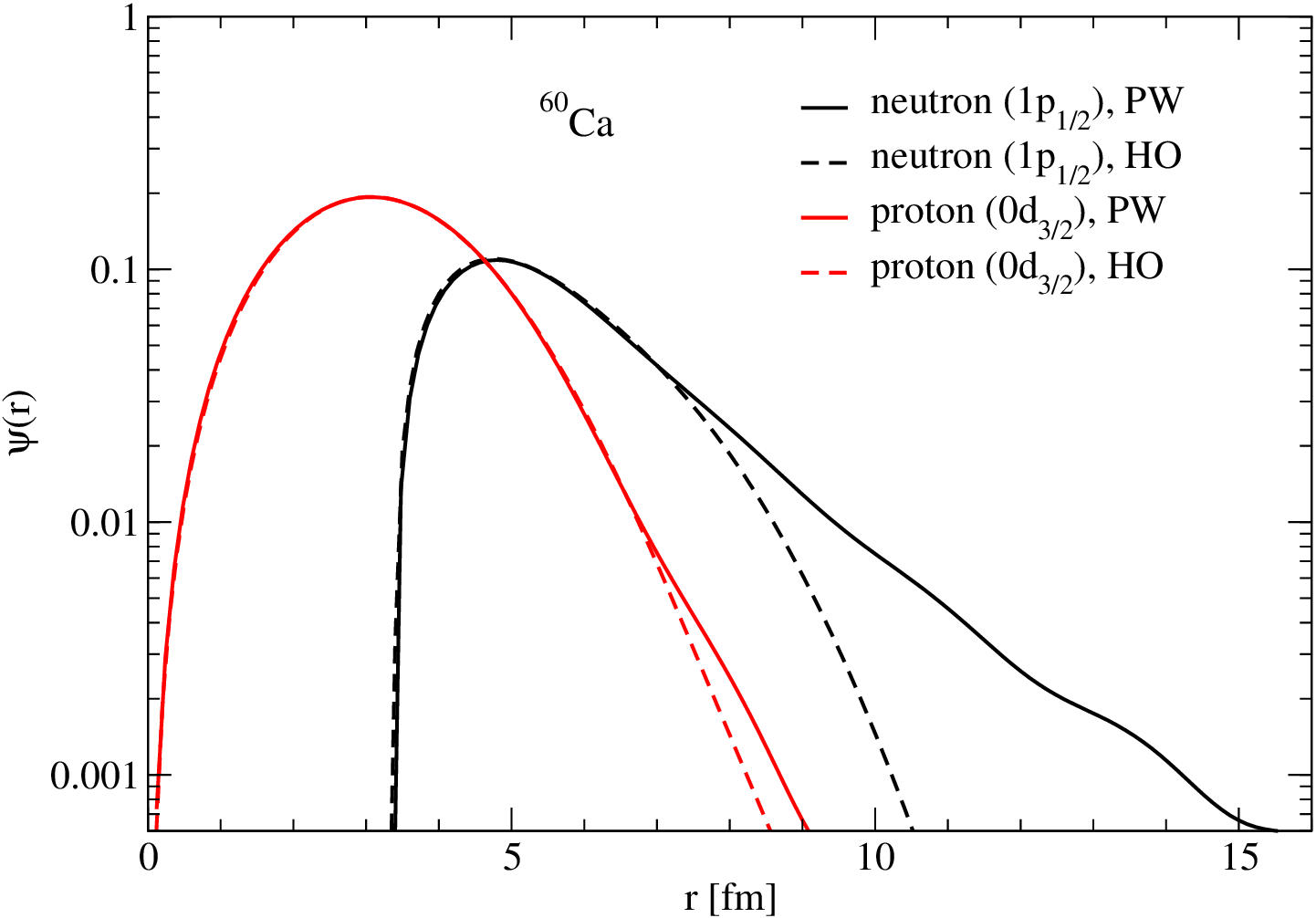}
\end{center}
\caption{(Color online) The neutron wave function of $1p_{1/2}$ (dark lines) and
the proton wave function of $d_{3/2}$ of $^{60}Ca$ (light lines) using two
different basis systems, the plane wave basis (solid lines) and the
harmonic oscillator basis (dashed lines).}
\label{fig:Ca60R}
\end{figure}

The differences in the asymptotic behavior of the wave functions are larger, if
we consider the shapes of the most weakly bound orbits for protons and neutrons
in the case of the weakly bound $^{60}Ca$, which are shown in 
Fig.~\ref{fig:Ca60R}. This is in line with our expectations that the PW basis is
superior to the HO basis in particular for the description of weakly bound
nucleons.  

\begin{figure}
\begin{center}
\includegraphics[width=0.6\textwidth]{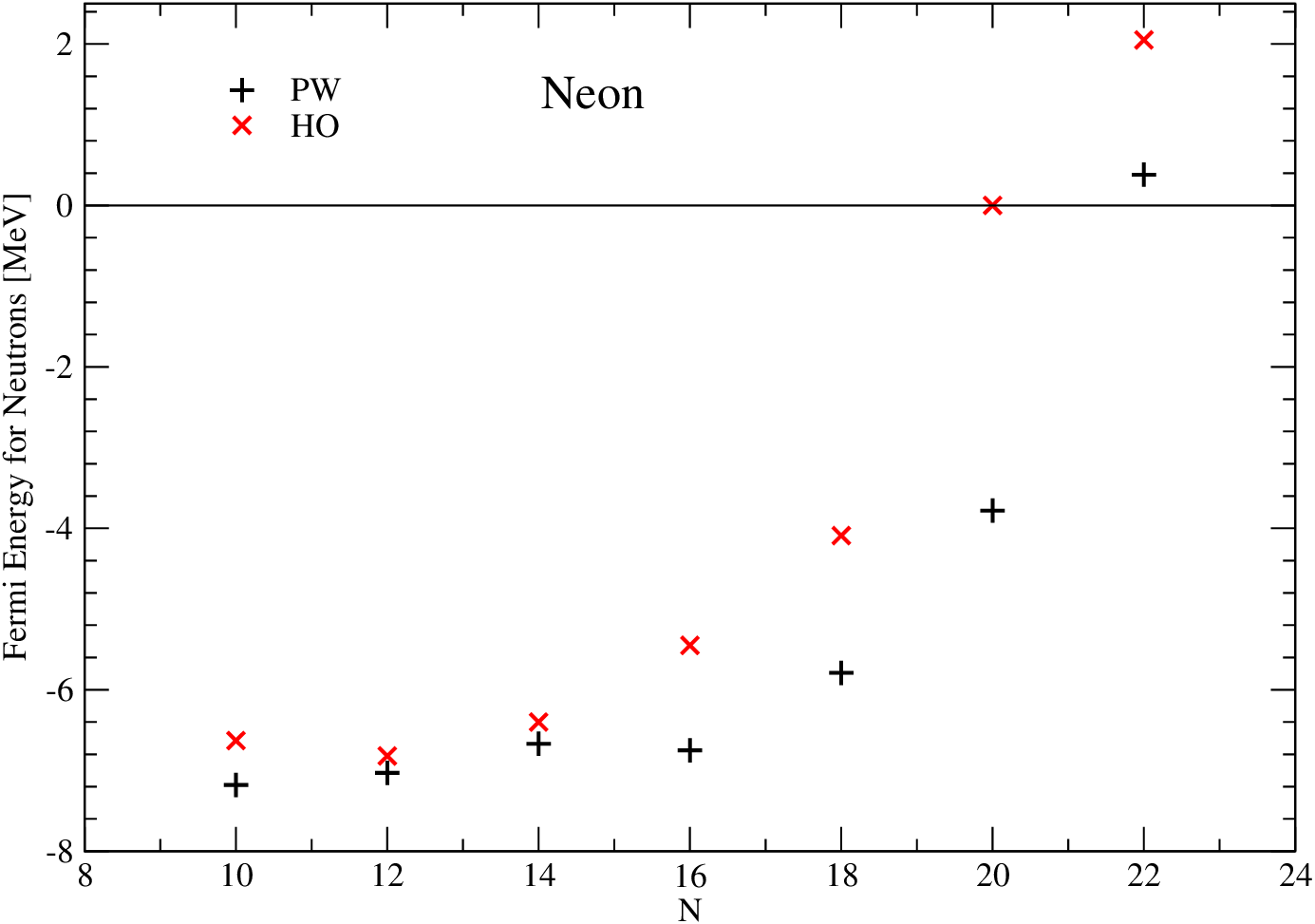}
\end{center}
\caption{(Color online) The Fermi energies for the neutrons in the isotopes of
$Ne$. Results of HF plus BCS calculations using the plane wave basis (PW) and
the harmonic oscillator basis (HO) are shown for various even numbers of
neutrons. }
\label{fig:ne20fn}
\end{figure}

Finally, we want to study the effect of the basis to be used on the  predictions
for the occurrence of the neutron drip line. For that purpose we perform HF plus
BCS calculations, as described in section \ref{sec:Vlowk_mat_el}, using the HO
as well as the PW basis. As an example we present the results for Fermi energy
of neutrons for the $Ne$ isotopes ranging from $A=20$ to $A=32$ in
Fig.~\ref{fig:ne20fn}. One can see that the differences in the predictions are
rather small for the tightly bound nuclei up to $^{24}Ne$ and is getting larger
with increasing number of neutrons. In this example the oscillator model would
predict stable isotopes up to $^{28}Ne$, while the calculation using the PW
basis predicts stable isotopes up to $^{30}Ne$.

\section{Summary and Conclusion}
\label{sec:SC}

It has been the aim of this study to establish a tool, which allows a
microscopic description of nuclei closed to the neutron drip and nuclear
structures in the transition from isolated nuclei to homogeneous matter based on
a realistic NN interaction.  A model space technique, the unitary-model-operator
approach (UMOA) \cite{suzuki:1982}, has been used to  disentangle the
low-momentum parts from the high-momentum parts from the  CD Bonn
interaction\cite{mach1}. This leads to a universal effective interaction
$V_{lowk}$, which is essentially independent on the underlying realistic
interaction $V$, if the cutoff $\Lambda$ is chosen around $\Lambda$ = 2
fm$^{-1}$.

The Hartree-Fock (HF) calculations using $V_{lowk}$ do not give any saturation
in nuclear matter and also HF calculation for finite nuclei lead to nuclear
systems of high density and large binding energies. It has been suggested to
compensate this feature by adding three-body forces~\cite{bogner:2005}. Instead
we supplement the $V_{lowk}$ interaction by a density-dependent contact
interaction and adjust the parameters in such a way that HF calculations for
infinite matter reproduce the empirical saturation point for symmetric matter
and the symmetry energy. 

This interaction model is than used to evaluate the properties of finite nuclei.
For nuclei with closed shells for protons and neutrons we obtain a very good 
agreement with the empirical data. Special attention is paid to the technique
for solving the HF equations for finite nuclei. The standard technique,
expanding the single-particle wave functions in a basis of appropriate harmonic
oscillator states is compared to a technique in which these states are expanded
in a basis of plane waves. Rather close agreement between these approximation
schemes is obtained for nuclei, which are tightly bound. The plane wave basis is
favorable, however, for calculating loosely bound systems, since it allows for
a better description of the asymptotic part of the nucleon wave functions.

These tools, the interaction model and the technique for solving the HF
equations, can now be used to describe the structure of loosely bound nuclei,
like halo nuclei or nuclei at the edges of the valley of stability, or the
nuclear structures in the so-called pasta phase of neutron stars. Another
perspective is to consider a many-body approach for describing the nuclear
systems, which goes beyond the HF approximation.

\section{Acknowledgements}
This work has been supported by a grant (Mu 705/5-1) of the Deutsche 
Forschungsgemeinschaft DFG


\begin{thebibliography}{99}

\bibitem{sk1} T.H.R. Skyrme, Nucl. Phys. {\bf 9}, 615 (1959).

\bibitem{sk2} D. Vautherin and D.M. Brink, Phys. Rev. C{\bf 5}, 626, (1972).

\bibitem{bv81} P. Bonche and D. Vautherin, Nucl Phys. A{\bf 372}, 496 (1981).

\bibitem{Oyamatsu:2007}{K. Oyamatsu and K. Iida, Phys. Rev. {\bf C75},  015801
(2007).}

\bibitem{Montani:2004}{F. Montani, C. May, and H. M\"{u}ther, Phys. Rev. C {\bf
69}, 065801 (2004).}

\bibitem{Goegelein:2007}{P. G\"{o}gelein and H. M\"{u}ther, Phys. Rev. C{\bf 77},
024312 (2008).}

\bibitem{goegeleina:2008}{P. G\"ogelein, E. N. E. van Dalen, C. Fuchs, and H.
M\"uther, Phys. Rev. C \textbf{77}, 025802 (2008).}

\bibitem{mach1} R. Machleidt, F. Sammarruca, and Y. Song, Phys. Rev. C\textbf{ 53},
R1483 (1996).

\bibitem{arg1} R.B. Wiringa, V.G.J. Stoks, and R. Schiavilla, Phys. Rev. C
\textbf{51}, 38 (1995).

\bibitem{Muether00}{ H. M\"uther and A. Polls, Prog. Part. Nucl. Phys. {\bf 45},
	243 (2000).}

\bibitem{Pieper:2006} S.C. Pieper, R.B. Wiringa, and J. Carlson, Phys. Rev. C
\textbf{70}, 054325 (2006). 	

\bibitem{bogner01} S. K. Bogner, T.T.S. Kuo, and L. Coraggio, Nucl. Phys.
\textbf{A684}, 432c (2001).

\bibitem{bogner:2005}{S.K. Bogner, A. Schwenk, R.J. Furnstahl, and A. Nogga, Nucl.
Phys. \textbf{A763}, 59 (2005).}

\bibitem{bogner:2007}{S. K. Bogner, R. J. Furnstahl, S. Ramanan, and A. Schwenk,
Nucl. Phys. \textbf{A784}, 79 (2007).}

\bibitem{bozek:2006}{P. Bo\.{z}ek, D.J. Dean, and H. M\"uther, Phys. Rev. C
\textbf{74}, 014303 (2006).}

\bibitem{kuckei:2003} J. Kuckei, F. Montani, H. M\"uther, and A. Sedrakian, Nucl.
Phys. \textbf {A 723}, 32 (2003).

\bibitem{coraggio:2005}{L. Coraggio, A. Covello, A. Gargano, N. Itaco, T. T. S.
Kuo, and R. Machleidt, Phys. Rev. C \textbf{71}, 014307 (2005).}

\bibitem{coraggio:2006}{L. Coraggio, A. Covello, A. Gargano, N. Itaco, and T. T. S.
Kuo, Phys. Rev. C \textbf{75}, 057303 (2007); Phys. Rev. C \textbf{73}, 014304
(2006).}

\bibitem{suzuki:1982}{K. Suzuki, Prog. Theoret. Phys. \textbf{68}, 246 (1986).}

\bibitem{fuji:2004} S. Fujii, R. Okamoto, and K. Suzuki, Phys. Rev. C \textbf{69},
034328 (2004).

\bibitem{moshinsky:1959}{M. Moshinsky, Nucl. Phys. \textbf{13}, 104 (1959).}

\bibitem{talmi:1952} {I. Talmi, Helv. Phys. Acta \textbf{25}, 185 (1952).}

\bibitem{talmi:1993} {I. Talmi, ``Simple Models of Complex Nuclei'', Harwood
Academic Publishers (Chur, Switzerland, 1993).}

\bibitem{Kung:1979} {C.L. Kung, T.T.S. Kuo, and K.F. Ratcliff, Phys. Rev. C
\textbf{19}, 1063 (1979).}

\bibitem{bonatsos:1989} {D. Bonatsos and H. M\"uther, Nucl. Phys. \textbf{A496}, 
23 (1989).}

\bibitem{DP:Bertsch91}  G.F. Bertsch and H. Esbensen, Ann. Phys. {\bf 209}, 327
(1991).

\bibitem{PP:Garrido99}  E. Garrido, P. Sarriguren, E. Moya de Guerra,
and P. Schuck,  Phys. Rev. C {\bf 60}, 064312 (1999).

\bibitem{coester:1970} F. Coester, S. Cohen, B.D. Day, and C.M. Vincent, Phys. Rev. C
\textbf{1}, 769 (1970). 

\bibitem{lejeune:2000} A. Lejeune, U. Lombardo, and W. Zuo,  Phys.Lett. B
\textbf{477}, 45 (2000).

\bibitem{brock:1984} R. Brockmann and R. Machleidt, Phys. Lett. B {\bf
149}, 283 (1984).

\bibitem{muemach:1990} H. M\"uther, R. Machleidt, and R. Brockmann, 
Phys. Rev. C{\bf 42}, 1981 (1990).

\bibitem{vandalen:2004b} E. N. E. van Dalen, C. Fuchs, and Amand Faessler, Nucl.
Phys. \textbf{A744}, 227 (2004).

\bibitem{vandalen:2007}{E.N.E. van Dalen, C. Fuchs, and  A. Faessler, Eur. Phys. J.
A \textbf{31}, 29 (2007).}

\bibitem{stoicea:2004} G. Stoicea {\it et al.} [FOPI Coll.], Phys. Rev. Lett. {\bf 92},
072303 (2004).

\bibitem{sturm:2001} C. Sturm {\it et al.} [KaoS Coll.], Phys. Rev. Lett. {\bf 86},
39 (2001); C. Fuchs, A. Faessler, E. Zabrodin, Y.M. Zheng, Phys. Rev. Lett. {\bf
86}, 1974 (2001); C. Fuchs, Prog. Part. Nucl. Phys. {\bf 56}, 1 (2006).

\bibitem{coraggio:2003}{L. Coraggio, N. Itaco, A. Covello, A. Gargano, and T. T.
S. Kuo, Phys. Rev. C \textbf{68}, 034320 (2003).}

\bibitem{hofmann:2001}{F. Hofmann, C.M. Keil, and H. Lenske, Phys. Rev. C
\textbf{64}, 034314  (2001).}

\end{thebibliography}
\end{document}